\documentclass[a4paper,twocolumn,prl,superscriptaddress,amsmath,amssymb,
               mathrsfs,showpacs]{revtex4-1}

\usepackage{hyperref}
\hypersetup{%
    colorlinks,
    linkcolor={black},
    citecolor={blue},
    urlcolor={black}}

\usepackage{braket}
\usepackage{epstopdf}
\usepackage{graphicx}
\usepackage{mathtools} 
\usepackage{dsfont} 
\usepackage{ifthen}
\usepackage{units}
\usepackage{xr} 

\newcommand{\ssec}[1]{\emph{#1} ---}
\newcommand{\figref}[1]{Fig.~\ref{#1}}
\renewcommand{\eqref}[1]{Eq.~(\ref{#1})}

\newcommand{\op}[2][]{\hat{#2}     \ifthenelse{\equal{#1}{}}{}{_\text{#1}}}
\newcommand{\scal}[2][]{#2         \ifthenelse{\equal{#1}{}}{}{_\text{#1}}}
\DeclareMathOperator{\Tr}{Tr}
\DeclarePairedDelimiter\abs{\lvert}{\rvert}
\newcommand{\com}[2]{\left[ #1 ,\, #2 \, \right]}
\newcommand{\acom}[2]{\left\{ #1 ,\hspace{2pt} #2 \right\}}
\newcommand{\ho}{s} 
\newcommand{\ra}{m} 
\newcommand{\cs}{S} 
\newcommand{\one}{\op{\ensuremath{\mathds{1}}}}

\begin{document}

\title{Quantum simulation of non-Markovianity using the quantum Zeno effect}

\author{Sabrina Patsch}
\affiliation{Theoretische Physik, Universit\"{a}t Kassel,
Heinrich-Plett-Stra{\ss}e 40, D-34132 Kassel, Germany}

\author{Sabrina Maniscalco}
\affiliation{%
  QTF Centre of Excellence, Turku Centre for Quantum Physics, Department of
  Physics and Astronomy, University of Turku, FIN-20014 Turku, Finland
}
\affiliation{%
  QTF Centre of Excellence, Department of Applied Physics, Aalto University,
  FIN-00076 Aalto, Finland
}

\author{Christiane P. Koch}
\affiliation{Theoretische Physik, Universit\"{a}t Kassel,
Heinrich-Plett-Stra{\ss}e 40, D-34132 Kassel, Germany}
\email{christiane.koch@uni-kassel.de}

\date{\today}

\begin{abstract}
  We suggest a quantum simulator that allows to study the role of memory effects
  in the dynamics of open quantum systems.  Our proposal is based on a bipartite
  quantum system consisting, for simplicity, of a harmonic oscillator and
  a few-level system; it exploits the formal analogy between dissipation and
  quantum measurements. The interaction between the subsystems gives rise to
  quantum Zeno dynamics, and the dissipation strength experienced by the
  harmonic oscillator can be tuned by changing the  parameters of the
  measurement, i.e., the interaction with the few-level system. Extension of the
  proposal to anharmonic systems is straightforward.
\end{abstract}

\maketitle

Quantum simulation uses a controllable quantum system to study another less
controllable quantum system ~\cite{GeorgescuRMP14}. It promises to advance the
study of many-body dynamics in condensed matter physics~\cite{GeorgescuRMP14},
but application to open quantum systems is also most
natural~\cite{Maniscalco2005, BarreiroNat11, LiuNatPhys11, MostameNJP12,
ChiuriSciRep12, SwekePRA14, SwekePRA15, SwekePRA16, BernardesSciRep16,
AlvarezPRA17, ChenuPRL17, GoviaQST17, MascarenhasPRA17, ChenPRL18, LiuNatComm18,
CuevasSciRep19}.  Quantum simulation of open quantum system dynamics can be
based on algorithmic
methods~\cite{SwekePRA14,SwekePRA15,SwekePRA16,AlvarezPRA17}, stochastic
Hamiltonians~\cite{ChenuPRL17} and Hamiltonian ensembles~\cite{ChenPRL18}, or
stochastic quantum walks~\cite{GoviaQST17}.  Experimental realizations to date
have focussed on qubit dynamics, encoding the system in trapped
ions~\cite{BarreiroNat11}, photon
polarization~\cite{LiuNatPhys11,ChiuriSciRep12,CuevasSciRep19,LiuNatComm18} and
nuclear spins~\cite{BernardesSciRep16}, but more complex systems such as the
light harvesting complex have also been suggested~\cite{MostameNJP12}.   A key
interest in the quantum simulation of open quantum systems is the controllable
transition from Markovian to non-Markovian dynamics. It implies the ability to
study memory effects and is motivated in a two-fold way. First, non-Markovian
dynamics are ubiquitous in condensed phase~\cite{WeissBook}, encountered in
settings as different as light harvesting or solid-state based quantum
technologies, but  inherently difficult to describe and
study~\cite{deVegaRMP17}. Second, memory effects may present a resource that can
be exploited~\cite{Wu2019,Anand2019}. For example, non-Markovian dynamics were
shown to allow for more efficient cooling~\cite{SchmidtPRL11}, better quantum
communication~\cite{LaineSciRep14,BylickaSciRep14}, and improved quantum gate
operation~\cite{RebentrostPRL09,ReichSciRep15}. The different physical aspects
underlying these improvements are captured by the various measures to quantify
non-Markovianity~\cite{RivasRPP14,BreuerRMP16}.  The transition from Markovian
to non-Markovian dynamics is comparatively easy to achieve when system and
environmental degrees are carried by the same physical object, for example the
polarization and frequency of a photon~\cite{LiuNatPhys11,LiuNatComm18}.
However, for a multi-partite implementation of the open system simulator one
needs to engineer suitable environmental correlations or a  sufficiently strong
interaction between system and environment for memory effects to come into play.
The latter has been realized by implementing the simulated interaction between
system and environment by quantum circuits~\cite{BernardesSciRep16}, whereas the
former is possible when using controllable systems to represent the environment,
such as quantum inductor-resistor-capacitor oscillators in a superconducting
qubit architecture~\cite{MostameNJP12}.

\begin{figure}[b]
  \includegraphics{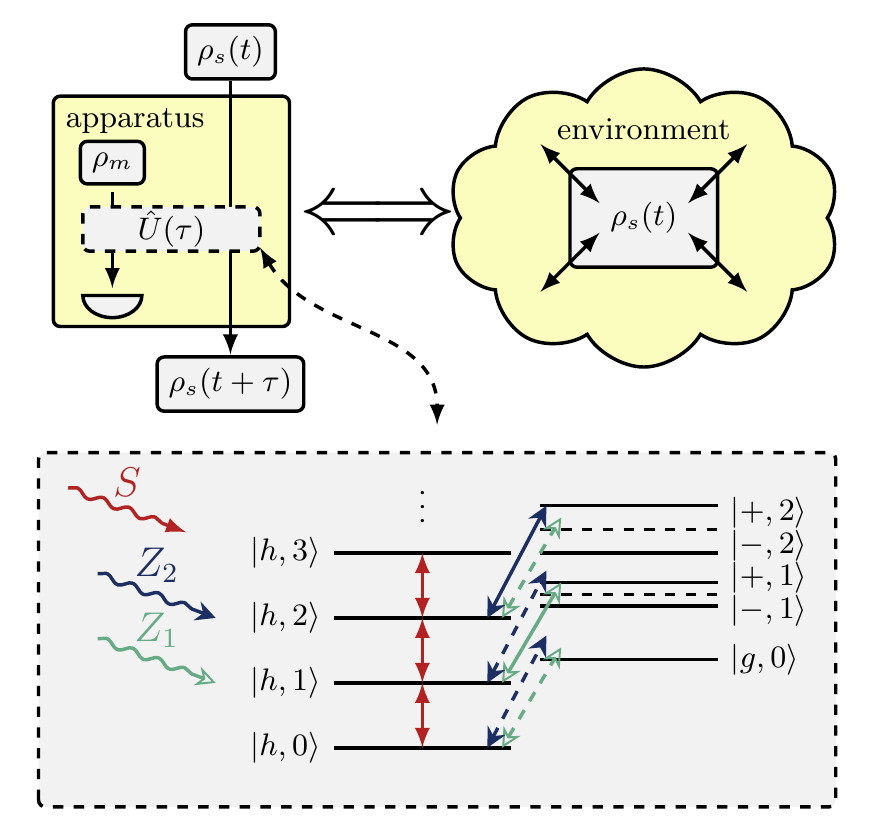}
  \caption{%
    (Color online) Analogy between the indirect measurement of a system
    $\rho_\ho$ using a meter $\rho_\ra$ (top left) and an open quantum system
    (top right). Bottom: Energy level diagram of a bipartite system consisting
    of a harmonic oscillator and a few-level system, interacting with three
    external fields---the drive $S$ of the harmonic oscillator (red), and two
    Zeno pulses $Z_{z=1}$ (green, open notched arrowhead) and $Z_{z=2}$ (blue,
    notched arrowhead) addressing the levels $\ket{z=1}$ and $\ket{z=2}$.
    Arrows with solid (dashed) lines indicate resonant (off-resonant)
    transitions.
  }
  \label{fig:setup}
\end{figure}
Here, we provide a general framework for inducing and tailoring essentially
arbitrary open system dynamics based on quantum measurements.  To this end, the
system is coupled to a meter which is measured destructively after a short
interaction time, cf.  \figref{fig:setup} (top). In the limit of
quasi-continuous measurements, this leads to confinement of the systems dynamics
to a specific subspace, i.e., to quantum Zeno dynamics~\cite{Raimond2012}.  We
show here that the control available in the measurement process can be used to
tune the system dynamics from being memory-less to exhibiting effects of memory.

\ssec{Theoretical framework}
For simplicity, we take the system that is connected to a tunable source of
dissipation to be a harmonic oscillator (HO), $\op[\ho]{H} = \sum_n \omega
n \ket{n}\bra{n}$, where we have set the vacuum energy to zero ($\hbar = 1$
throughout).  It would also be possible to assume a finite $n$-level system
instead for achieving qualitatively the same results.  The system is coupled to
a meter which we model by a three-level system with states $\ket{g}$, $\ket{e}$,
and $\ket{h}$, $\op[\ra]{H} = \omega' \ket{g}\bra{g} + \big( \omega
+ \omega'\big) \ket{e}\bra{e}$. The energy of $\ket{h}$ is set to zero and the
$\ket{g}\leftrightarrow\ket{e}$ transition is resonant with the HO transition
frequency $\omega$ while the $\ket{h}\leftrightarrow\ket{g}$ transition is far
off-resonant with $\omega \ll \omega'$. The meter can easily consist of more
than three levels but existence of one resonant and at least one off-resonant
transition is essential.  The resulting coupling, in the interaction picture
with respect to the drift Hamiltonian, $\op[0]{H} = \op[\ho]{H} + \op[\ra]{H}$,
and using the rotating wave approximation, can then be described by
a Jaynes-Cummings type interaction~\cite{Haroche2006},
\begin{align}
  \op[\ho\ra]{H} = \frac{\Omega}{2}\left(\op[-]{\sigma}\op{a}^\dagger
                 + \op[+]{\sigma}\op{a} \right)
  \label{eq:H_JC}
\end{align}
with Rabi frequency $\Omega$, $\op[-]{\sigma}=\ket{g}\bra{e}
= \op[+]{\sigma}^\dagger$, and creation and annihilation operators
$\op{a}^{(\dagger)}$ for the HO.  The eigenstates of $\op[\ho\ra]{H}$ are the
dressed states $\ket{g,0}$, and $\ket{\pm,n} = \frac{1}{\sqrt{2}} \left(
\ket{e,n-1} \pm \ket{g,n} \right)$ for $n \ge 1$ with eigenenergies $E_n^{\pm}=
\omega n \pm \frac{\Omega}{2}\sqrt{n}$~\cite{Haroche2006}, cf.\
\figref{fig:setup} (bottom).  In addition to its coupling to the meter, the HO
is driven by a resonant classical source $S$ with field strength $\alpha$,
\begin{align}
  \op[\cs]{H} = \alpha \op{a}^\dagger + \alpha^* \op{a}\,.
  \label{eq:H_CS}
\end{align}
We assume the drive to be weak compared to the coupling with the meter, $\alpha
\ll \Omega$, such that the HO will only be driven if the meter is in $\ket{h}$.
In this limit, $\op[\cs]{H}$ will induce a displacement $\beta = -i \alpha \tau$
of the HO's state , $\op[\cs]{U}(\tau) = e^{-i\op[\cs]{H} \tau} = e^{\beta
\op{a}^\dagger - \beta^* \op{a}}$.

A tunable source of dissipation is introduced by a series of indirect
measurements of the HO's state using identical meters, cf.  \figref{fig:setup}
(top). These measurements shall determine  whether the HO is in a specific Fock
state $\ket{z}$, the 'Zeno level'.  To this end,  HO and  meter
are coupled by a so-called Zeno pulse $Z_z$ with  coupling strength $\omega_z$
which is resonant to the $\ket{h,z}\leftrightarrow\ket{+,z}$ transition,
\begin{align}
  \op{H}_{z} = \frac{\omega_z}{2}
  \Big( \ket{h,z}\bra{+,z}+ \ket{+,z}\bra{h,z} \Big)\,,
  \label{eq:H_z}
\end{align}
cf. \figref{fig:setup} (bottom).  If and only if the HO is in the Zeno level,
$Z_{z}$ will induce Rabi oscillations, $\op[z]{U}(\tau) = e^{-i\op[z]{H} \tau}
= e^{\left(\ket{h,z}\bra{+,z}+ \ket{+,z}\bra{h,z} \right)\phi_z/2}$,
accumulating a Rabi angle $\phi_z = \omega_z \tau$ during the interaction time
$\tau$.  Thus, there will only be population in the state $\ket{+,z}$ if there
was population in $\ket{z}$ initially. A subsequent destructive measurement of
the meter at time $\tau$, corresponding to a partial trace over the meter, thus
provides information on the population of $\ket{z}$~\cite{Raimond2012}.  The
state of the HO at the end of one time interval $\tau$ can be obtained in terms
of piecewise dynamics,
\begin{align}
  \op[\ho]{\rho}(t+\tau)
  = \scal[\ra]{\Tr} \left\{
    \op{U}(\tau)\op{\rho}(t)\op{U}^{\dagger}(\tau)
    \right\}
  \label{eq:piecewise_dyn}
\end{align}
with $\op{U}(\tau) = e^{-i\left(\op[\cs]{H} + \op[\ho\ra]{H}
+ \op[Z]{H}\right)\tau}$ and  $\op{\rho}$ $\left(\op[\ho]{\rho}\right)$ denoting
the density operator of the bipartite system (the HO). The procedure of coupling
the HO to a meter during time $\tau$ and performing a destructive measurement of
the meter afterwards is repeated several times, with the new meter initially
always in $\ket{h}$. For small displacements, $\abs{\beta} \ll 1$,   of the HO
in between two measurements, the protocol gives rise to quantum Zeno dynamics
(QZD)~\cite{Raimond2012,Facchi2002,Facchi2008}.  Hence, when choosing an initial
state of the system below the Zeno level, $\ket{n_0 < z}$, the dynamics is
confined to the Zeno subspace $\mathcal{H}_{Z}^{<} = \{ \ket{0}, \dots,
\ket{z-1} \}$.  The displacement $\beta$, due to the drive $S$, and the Rabi
angle $\phi_z$, due to the Zeno pulse $Z_z$, act as ``knobs'' to control the
dynamics of the quantum simulator as we show below.

The proposed scheme can be realized experimentally in several ways.  In cavity
or circuit QED, the HO corresponds to a cavity mode while three neighboring
circular states of a Rydberg atom flying through the cavity~\cite{Raimond2012}
or a superconducting qubit coupled to the cavity~\cite{Blais2004} encode the
three-level system. The example of a cavity QED implementation is discussed in
more detail in the Supplemental Material (SM)~\cite{SM}.  Similarly, a common
vibrational mode of trapped ions together with ion qubit states also realize the
envisioned bipartite system~\cite{Kienzler2015}.  However, the system does not
need to be harmonic for the model to work.  The model is even more general since
the system does not have to be bipartite---it is sufficient to identify two
different subspaces in the quantum system: a `system' and a `meter' part and
carry out a state-selective measurement of the `system' using the `meter'.  An
example for a unary realization would be a Bose-Einstein condensate (BEC) with
hyperfine ground state levels encoding the system~\cite{Schafer2014}.  In this
case, the open quantum system is matter-based, instead of photonic as in the
cavity or circuit QED setup.  In general, realization of our proposal requires
three conditions to be met:
(i) existence of  a `system' part whose dynamics are decoupled from the `meter'
part when driven by a source $S$,
(ii) selective excitation of system states by driving transitions in the meter
using the Zeno pulse $Z_z$,
(iii) subjecting meter states to dissipation.
Dissipation can be introduced by projective measurement of the meter, or it can
be natural such as a fast decay.  In the BEC example, the Zeno pulse drives
transitions to an excited state hyperfine manifold~\cite{Schafer2014} such that
the meter is subject to spontaneous decay.

\ssec{Simulating an open quantum system}
To illustrate the implementation of the simulator, we choose the overall
protocol duration such that $-i\alpha T = 2 \pi$, take the initial state to be
$\ket{n_0 = 0}$ and the Zeno level $\ket{z = 2}$ such that QZD occurs in the
two-dimensional Zeno subspace $\mathcal{H}_{Z} = \{ \ket{0}, \ket{1} \}$.  To
quantify the amount of dissipation, we use the linear entropy,  $S_L
= 1 - \Tr\{\op{\rho}^2\}$; and the population $P_{\bar{Z}}$ that has escaped
from the Zeno subspace $\mathcal{H}_Z$, $P_{\bar{Z}} = \sum_{n=2}^{\infty}
\braket{n|\op[\ho]{\rho}(T)|n}$, measures the QZD infidelity.

We expect the system dynamics to display memory effects and use the so-called
BLP measure quantifying non-Markovianity as accumulated revivals in
distinguishability of two initial states~\cite{Breuer2009}.  The optimal state
pair which maximizes the BLP measure for our choice of $\alpha$ and $\ket{z}$ is
given by
\begin{align}
  \ket{\Psi_1 (\vartheta)} =
  \cos\left( \tfrac{\vartheta}{2}\right) \ket{0}
  + \sin\left( \tfrac{\vartheta}{2}\right) \ket{1}
  ,\quad
  \ket{\Psi_2} = \ket{2}
  \label{eq:optimal_state_pair}
\end{align}
for all $\vartheta\in\left[0,2\pi\right)$~\cite{SM}, and we take $\vartheta=0$
in the following.

\begin{figure}[tb]
  \centering
  \includegraphics{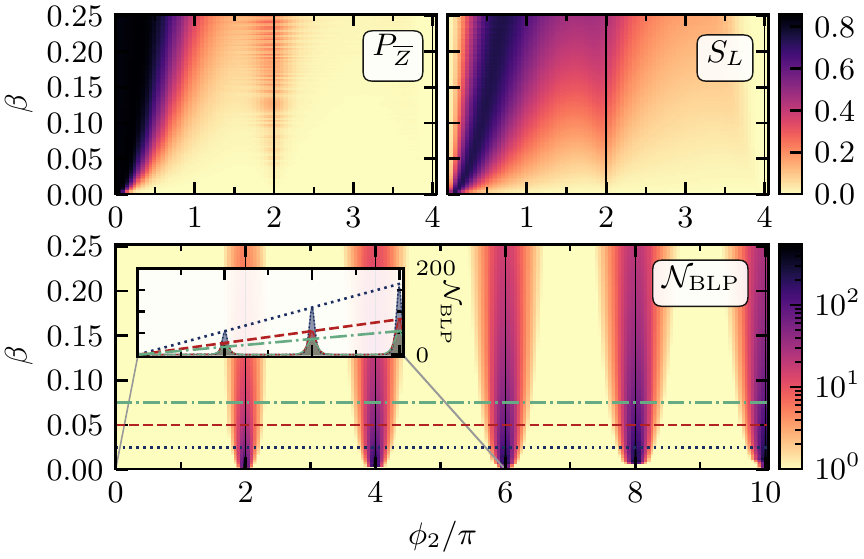}
  \caption{%
    (Color online)
    Infidelity $P_{\bar{Z}}$ of the quantum Zeno dynamics, dissipation $S_L$,
    and non-Markovianity measure $\mathcal{N}_\text{BLP}$ as a function of the
    displacement $\beta$ due to the classical drive $S$ and the Rabi angle
    $\phi_2$ accumulated due to the Zeno pulse $Z_2$. Inset:
    $\mathcal{N}_\text{BLP}$ as a function of $\phi_2$ for $\beta = 0.025$ (blue
    dotted, highest peaks), $0.050$ (red dashed) and $0.075$ (green dash-dotted,
    smallest peaks) with linearly increasing peak heights.
  }
  \label{fig:nonmarkov_colormap}
\end{figure}
We start by discussing the quantum simulator subject to the classical drive $S$
and only one Zeno pulse with Zeno level $\ket{z=2}$, cf.
\figref{fig:nonmarkov_colormap}. An arbitrary degree of dissipation, as
indicated by $S_L$, can be engineered by tuning the displacement $\beta$ induced
by the classical drive and the Rabi angle $\phi_{z=2}$ due to the Zeno pulse.
More specifically, large dissipation can be realized by choosing a small
$\phi_{2}$ while strong non-Markovianity can be engineered by setting  $\phi_2$
to multiples of $2\pi$.  Dissipation occurs mostly in the parameter range where
the population leaves the Zeno subspace \cite{SM}, except for $\phi_2
\rightarrow 0$, where no dissipation occurs since the coupling of system and
meter vanishes. Both the infidelity of the QZD and the dissipation are
vanishingly small for Rabi angles $\phi_2 > 4\pi$ \cite{SM} which is why this
parameter range is omitted from \figref{fig:nonmarkov_colormap}.

The signatures of non-Markovianity are strongest for Rabi angles $\phi_2
= 2n\pi$ with $n$ being an integer, cf. \figref{fig:nonmarkov_colormap}
(bottom).  This is because a pulse $Z_z$ with $\phi_z = 2n\pi$ corresponds to
the map $\ket{h,z} \mapsto (-1)^n\ket{h,z}$ after the time interval $\tau$,
i.e., the pulse changes only the phase of the Zeno level and no entanglement
between system and meter is left~\footnote{For small values of the displacement
$\beta$, this choice of $\phi_z$ results in a quantum non-demolition measurement
\cite{Nogues1999} preserving the information of the HO's state. For $\phi_z
= 2n\pi$ with odd $n$, corresponding to the map $\ket{h,z} \mapsto -\ket{h,z}$,
the dynamics are inverted by flipping the phase of the state similar to a spin
echo \cite{Hahn1950}.}.  For other Rabi angles, $\phi_z \neq 2n\pi$,
entanglement remains between system and meter at the end of the time interval
$\tau$. Measurement of the meter then erases information, resulting in less
distinguishable states in $\mathcal{N}_{BLP}$ and the system dynamics becoming
Markovian. In the Markovian limit, the dynamics can be described by a master
equation of Lindblad form by assuming quasi-continuous
measurements~\cite{SM},
\begin{widetext}
\begin{align}
    \frac{d\op[\ho]{\rho}(t)}{dt}
    = -i \com{\op[\cs]{H}}{\op[\ho]{\rho}(t)}
     + \kappa\gamma_A
        \left( \op{A}\op[\ho]{\rho}(t) \op{A}^\dagger
      - \frac{1}{2}\acom{\op{A}^\dagger\op{A}}{\op[\ho]{\rho}(t)} \right)
     + \kappa\gamma_\Pi
        \left( \op{\Pi}\op[\ho]{\rho}(t) \op{\Pi}^\dagger
      - \frac{1}{2}\acom{\op{\Pi}^\dagger\op{\Pi}}{\op[\ho]{\rho}(t)} \right)
    \label{eq:mastereq}
\end{align}
\end{widetext}
with $\kappa=1/\tau$, $\op{\Pi} = \ket{z}\bra{z}$ and $\op{A} = \ket{z-1}\bra{z}$, and
\begin{align}
  \gamma_A     = \frac{1}{2} \sin^2 \frac{\phi_z}{2},\,
  \gamma_\Pi^* = 2 \sin^2 \frac{\phi_z}{4}, \,
  \gamma_\Pi   = \left.\gamma_\Pi^*\right.^2 + \gamma_A.
  \label{eq:decay_rates}
\end{align}
Given the definitions of $\op A$ and $\op\Pi$,  $\kappa \gamma_A$ corresponds to
the rate of population transfer from the Zeno level to the level below, and
$\kappa \gamma_\Pi$ to the dephasing rate of the Zeno level. The master equation
provides yet another angle illustrating the functionality of the quantum
simulator, in addition to \figref{fig:nonmarkov_colormap}: The decay rates can
be varied by tuning the experimentally accessible parameters $\phi_z$ and
$\kappa$.  In the non-Markovian regime ($\phi_2 \simeq 2n\pi$) and for a given
displacement $\beta$, the value of the non-Markovianity measure increases
approximately linearly with the Rabi angle $\phi_2$ (see inset in
\figref{fig:nonmarkov_colormap}).  For constant $\phi_2$ in turn, the
non-Markovianity measure decreases as $\beta$ gets larger since it is inversely
proportional to the Zeno pulse field strength $\omega_Z = \phi_2/\tau \propto
\phi_2/\beta$.  However, with only a single Zeno pulse, dissipation and memory
effects cannot be tuned independently.

This can be remedied by employing two state-selective excitations
simultaneously, cf.~\figref{fig:setup}.  Specifically, we assume state-selective
excitation of the Fock state $\ket{2}$ by the Zeno pulse $Z_2$ and of $\ket{1}$
by another Zeno pulse $Z_1$ (for a system with a larger Hilbert space, any two
distinct levels can be chosen).  While $Z_2$ controls  memory effects, $Z_1$
induces dissipation. Large memory effects require a Rabi angle $\phi_2$ close to
multiples of $2\pi$, whereas small $\phi_1$, i.e., $\phi_1$ smaller than $\pi$
for our values of the displacement $\beta$, results in strong dissipation.  Note
that, despite employing two Zeno fields, still only a single measurement needs
to be carried out at the end of each time interval $\tau$.
\begin{figure}[tb]
  \centering
  \includegraphics{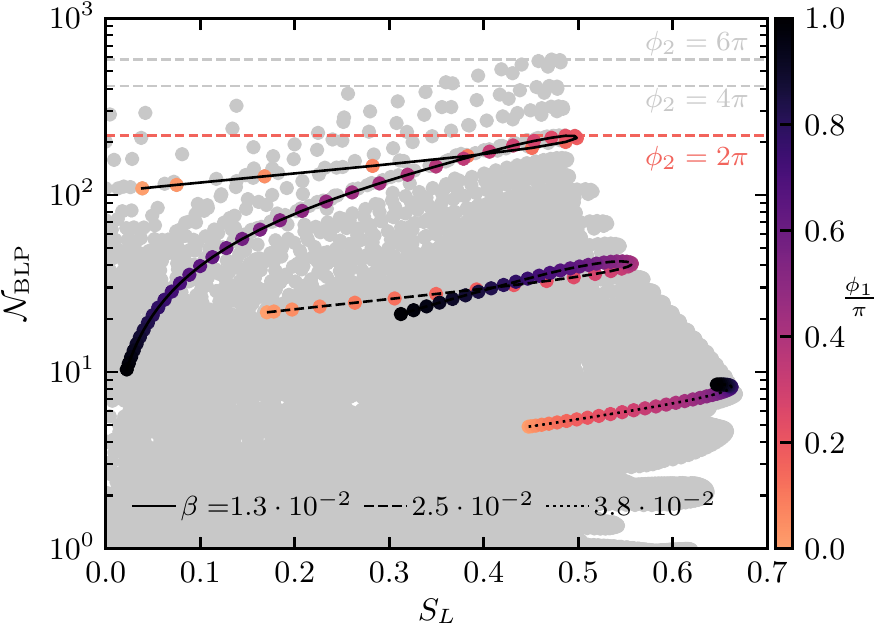}
  \caption{%
    (Color online)
    Attainable combinations of dissipation $S_L$ and non-Markovianity
    $\mathcal{N}_\text{BLP}$ using two Zeno pulses $Z_2$ and $Z_1$ (grey dots)
    with the colored dots highlighting the combinations for a Rabi angle $\phi_2
    = 2\pi$ and three different values of $\beta$. The color of the dots encodes
    the value of the Rabi angle $\phi_1$ as indicated by the color bar, and
    lines connect the dots with the same value of the displacement $\beta$ as
    given by the line style.  The horizontal dashed lines indicate the maximal
    value of $\mathcal{N}_\text{BLP}$ when tuning $\phi_2$ to $2\pi$, $4\pi$ and
    $6\pi$.
  }
  \label{fig:sim_ranges}
\end{figure}
\figref{fig:sim_ranges} shows that any combination of dissipation strength,
quantified by the linear entropy $S_L$, and memory effects, quantified by the
measure $\mathcal{N}_\text{BLP}$ (calculated with $\vartheta=\pi$~\cite{SM}),
can be attained by varying the displacement $\beta$ and the two Rabi angles
$\phi_2$ and $\phi_1$.  The gray dots in \figref{fig:sim_ranges} show all
combinations of $S_L$ and $\mathcal{N}_{BLP}$ obtained by sampling $\beta$ from
$1.26\cdot 10^{-2}$ to $0.251$ in $20$ steps, $\phi_2$ from $0$ to $6\pi$ in
steps of $0.1\pi$ and $\phi_1$ from $0$ to $\pi$ in steps of $0.025\pi$.  In
order to illustrate how the change in one of the three parameters affects the
dissipation strength and memory effects, we inspect the special case $\phi_2
= 2\pi$ for three different values of $\beta$, represented by different lines,
and varying $\phi_1$, as encoded in the color bar in \figref{fig:sim_ranges}.
When considering, for instance, the points on the solid line, corresponding to
the smallest considered value of $\beta$, an increase of the Rabi angle $\phi_1$
(moving from the lightest point on the upper left end to the right) leads to
a large increase of dissipation but initially only to a small increase degree in
the non-Markovianity measure.  When $\phi_1$ takes the value $0.175\pi$, the
line reaches a turning point where the dissipation takes its maximal value with
$S_L = 0.5$. A further increase in $\phi_1$ results in a decrease of  both
dissipation and non-Markovianity measure. This is because the pulse $Z_1$ starts
to build up a Zeno barrier at $\ket{1}$ which prevents dissipation.  For larger
$\beta$, cf. the dashed and dotted lines, the dissipation becomes stronger, and
a larger value of $\phi_1$ is needed to built up a Zeno barrier.  Accordingly,
the turning point of these lines is shifted to larger values of $S_L$ and
$\phi_1$. The picture is very similar for larger Rabi angles $\phi_2=2n\pi$ but
the maximally attainable degree of non-Markovianity increases, as indicated by
the dashed horizontal lines in \figref{fig:sim_ranges}. When $\phi_2$ takes
values inbetween multiples of $2\pi$, the non-Markovian measure decreases
quickly which allows to realize dynamics without memory effects. Since, in an
experiment, the three parameters $\beta$, $\phi_2$ and $\phi_1$ can be tuned
essentially continuously, the areas inbetween the gray dots can also be
attained.

While we have restricted our analysis here to a two-dimensional subspace of the
system's Hilbert space, generalization to larger subspaces is straightforward.
Indeed, to create a subspace with dimension $z_0$, one simply needs to choose
the Zeno level $\ket{z_0}$, i.e., adjust the frequency of the Zeno pulse
$Z_{z_0}$, accordingly.  The dimension  of the system should therefore at least
be three in order to enable non-trivial dynamics, but can be arbitrarily large
and even infinite as in the cavity QED application~\cite{Raimond2012}.
Moreover, it is also possible to generate dissipation and memory effects on
multiple system states $\ket{z_i}$ with $i\in\{1,2,\dots,N\}$ by varying the
frequency of the Zeno pulse $Z_{z_0}$ as a function of time or by employing
multiple Zeno pulses $Z_{z_i}$ and tailoring the Rabi angles as explained above.
The number $N$ of maximally controllable states is then limited by the dimension
of the `meter' subspace. In a cavity or circuit QED setup, this subspace is
formed by the dressed states and is thus infinite. In the BEC
experiment~\cite{Schafer2014}, the limit is given by the size of the second
hyperfine manifold.

\ssec{Conclusions}
We have introduced a quantum simulator for open quantum system dynamics based on
a series of indirect, state-selective measurements of a system by a meter.  The
driven system dynamics becomes open when tracing out the meter, i.e., our
proposal relies on the formal analogy between measurement and dissipation.  The
open system dynamics can be tuned from memory-less to displaying maximal memory
effects by choosing the amount of entanglement between system and meter at the
end of their interaction.  This represents a direct realization of memory
effects in terms of repeated interactions between system and environment and
past-future independence~\cite{Li2018}.  The amount of dissipation and the
degree of non-Markovianity can be engineered independently by a suitable choice
of the amplitudes of a classical drive for the system and two Zeno pulses
probing the system states dressed by the interaction with the meter, whereas the
frequency of the Zeno pulses determines which system levels are subject to
dissipation.  Our proposal opens the way to experimentally realize quantum
simulation of open quantum systems, studying memory effects in a controlled way.
This would allow, for example, to clarify the role of memory effects for the
controllability of an open quantum system~\cite{KochJPCM16} and address the
question whether and how memory effects alter control strategies for open
quantum systems.


\begin{acknowledgments}
  We thank J.-M. Raimond, S. Gleyzes, F. Ass\'{e}mat, and E. M. Laine for
  discussions.  Financial support from the European Union under the Research and
  Innovation action project ``RYSQ''(Project No. 640378), the DAAD/Academy of
  Finland mobility grants, and the Studienstiftung des deutschen Volkes e.V.\ is
  gratefully acknowledged. This work was partly done using the high-performance
  computing cluster FUCHS-CSC provided by the Center for Scientific Computing
  (CSC) of the Goethe University Frankfurt in the framework of the HHLR-GU
  (Hessisches Hochleistungsrechenzentrum der Goethe-Universit\"{a}t).
\end{acknowledgments}

\clearpage

\begin{appendix}

\section{Implementation of the quantum simulator in cavity QED}

The presented scheme is ideal for an experimental implementation in cavity QED
as proposed in Ref.~\cite{Raimond2012}. The harmonic oscillator can be
identified by a mode of the cavity and the three-levels of the meter can be
chosen to be three neighbouring circular states of the Rydberg atom,
$\ket{49\text{C}} \equiv \ket{h}$, $\ket{50\text{C}} \equiv \ket{g}$, and
$\ket{51\text{C}} \equiv \ket{e}$, flying through the cavity.  In a fountain
arrangement, the interaction time between cavity and  atom is sufficiently long
to perform a series of indirect measurements~\cite{Raimond2012}. Due to the
cryostatic environment and the long lifetime of circular states in Rydberg
atoms, `true' dissipation due to field energy damping or atomic relaxation is
negligible on the relevant timescale. The atom will be excited to the Rydberg
regime when it is already located inside the cavity which marks the beginning of
one sequence of the protocol.  In order to specifically address the
$\ket{h,z}\leftrightarrow\ket{+,z}$ transition using the Zeno pulse $Z_z$ and
distinguish it from the nearby $\ket{h,z}\leftrightarrow\ket{-,z}$ and
$\ket{h,z}\leftrightarrow\ket{-,z-1}$ transitions, the duration $\Delta t_{Z}$
of the pulse $Z$ has to be long enough, namely $\Delta t_{Z} \gg 1/\left(\Omega
\abs{\sqrt{z+1} - \sqrt{z}}\right)$ \cite{Raimond2012}. The end of one sequence
with interaction time $\tau$ is triggered by ionizing the atom within the cavity
by field ionization.

In the simulations presented in the main text, we take the transition
frequencies of the three-level system and the HO,
$\omega_{hg}=\omega'$ and $\omega_{ge}=\omega_\text{HO}=\omega$, and the Rabi
frequency $\Omega$ from Ref.~\cite{Raimond2012}.  The values of $\omega$ and
$\omega$ justify the assumption of $\omega \ll \omega'$.

\section{Identification of the optimal state pair in the non-Markovianity measure}

In order to quantify memory effects, we use the BLP measure~\cite{Breuer2009}.
Evaluation of this measure requires comparatively little numerical
effort---propagation of two well chosen initial states whereas the geometrical
measure for non-Markovianity given by the state space volume~\cite{Lorenzo2013},
for instance, requires to propagate a full set of basis states. The measure
proposed in \cite{Rivas2010} based on maximally entangling the system with an
ancillary system leads to a further increase of the total dimension of the
Hilbert space and is hence not practical for our purposes either.  An obstacle
for using the BLP measure is that it requires optimization over the system
Hilbert space in order to identify the optimal initial state pair. On first
glance, this seems difficult for the infinitely large Hilbert space of
a harmonic oscillator. Nonetheless, the BLP measure turns out to be suitable in
our case since we can reduce the size of the subspace in which we have to
perform the optimization significantly by using our knowledge about the tailored
interaction of the HO and the meter.

\begin{figure}
  \centering
  \includegraphics{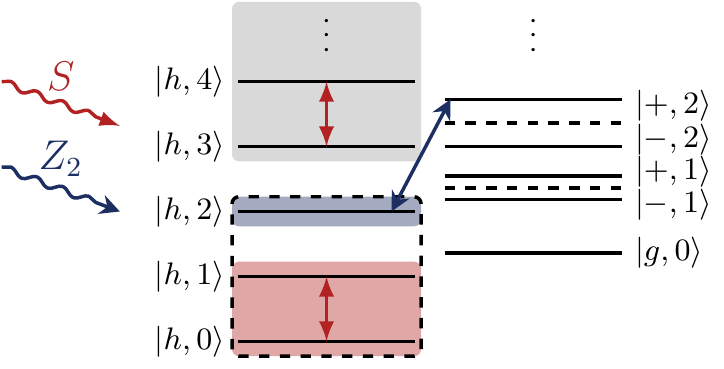}
  \caption{%
    Energy level diagram of the bipartite system indicating different
    subspaces: The Zeno subspace $\mathcal{H}_{Z}$ (red shaded), the Zeno level
    $\ket{z}$ (blue shaded), the extended Zeno subspace $\mathcal{H}_{Z}^+$
    (black dashed box) and the subspace with purely unitary dynamics (grey
    shaded).
  }
  \label{sfig:subspaces}
\end{figure}

The reduction of the system size goes as follows, cf.~\figref{sfig:subspaces}.
We know that the Zeno pulse only affects the Zeno level and the level below
since the dressed state we couple to is defined as $\ket{+,z} = \tfrac{1}{\sqrt
2}(\ket{e,z-1} + \ket{g,z})$. If the initial state is located within the Zeno
subspace $\mathcal{H}_{Z}$, signatures of non-Markovianity can only arise from
population in states within the extended Zeno subspace $\mathcal{H}_{Z}^+ = \{
  \ket{0}, \dots, \ket{z} \}$,  whereas the dynamics in the remaining Hilbert
space are purely unitary. Numerical tests confirm this conjecture. It is thus
sufficient to consider only the extended Zeno subspace for the optimization. As
in the main paper, we seek to find the optimal state pair for the special case
of $z=2$. Here, the extended Zeno subspace is only three-dimensional.

The optimization can be further simplified by considering properties of optimal
state pairs for the BLP measure in general. To calculate the BLP measure,
revivals of distinguishability of two initial states during a fixed time are
accumulated. The distinguishability is calculated using the trace distance of
the state pair at a given time. From this we can conclude that the trace
distance of the optimal state pair shows revivals for the longest time with
a maximal peak amplitude as compared to all other state pairs. Ideally, the
states oscillate between being fully distinguishable, with trace distance equal
to $1$, and fully indistinguishable, with trace distance equal to $0$. Thus, the
dynamics of the two states which form the optimal pair should be as different
from each other as possible. In our model, the Zeno level and a state within the
Zeno subspace form such a pair. The former is subject to the quantum Zeno effect
which means that the population in the Zeno level is frozen. For the latter,
quantum Zeno dynamics (QZD) are induced and the population never leaves the Zeno
subspace. Thus, in an ideal Zeno situation, this pair of initial states stays
distinguishable forever. However, in our realistic model with time-resolved
dynamics, the Zeno pulse introduces Rabi oscillations in the bipartite system.
Thus, the overlap between the two states varies with time, leading to
oscillations of the trace distance which indicate information flow and
non-Markovianity. More explicitly, every time the Zeno pulse $Z_2$ induces
a $\pi$-pulse on the bipartite system, half of the population that has been in
the Zeno level, $\ket{h,2}$, is driven towards the level below, $\ket{e,1}$.
Thus, the distinguishability from a state within the Zeno subspace decreases. If
$\phi_2>\pi$, the population oscillates back to the state $\ket{h,2}$, such the
distinguishability increases again and the trace distance undergoes a revival.
To summarize, the optimal state pair consists of the Zeno level on the one hand
and a state from the Zeno subspace on the other hand.

\begin{figure}
  \centering
  \includegraphics[trim=0 0 0 1cm,clip]{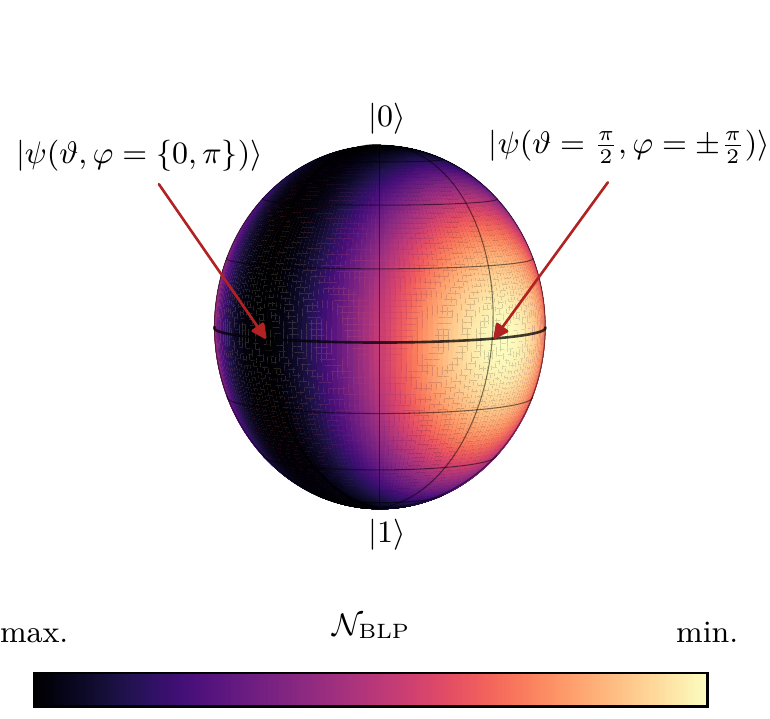}
  \caption{%
    BLP measure $\mathcal{N}_\text{BLP}$ for different initial states
    $\ket{\Psi_1} = \ket{\psi (\vartheta,\varphi)}$ on the Bloch sphere spanned
    by the Zeno subspace $\{ \ket{0}, \ket{1}\}$. The second state of the state
    pair is the Zeno level $\ket{\Psi_2} = \ket{2}$ outside of this subspace.
    The parameters are $\beta = 0.025$ and $\phi_2=4\pi$.
  }
  \label{sfig:optimal_state_pair}
\end{figure}

\begin{figure}
  \centering
  \includegraphics{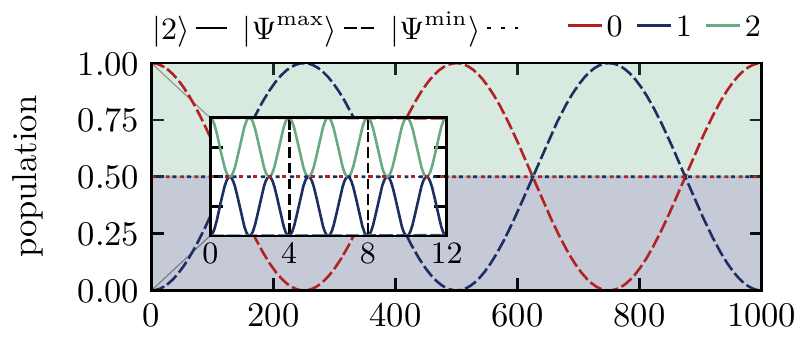}\\
  \includegraphics{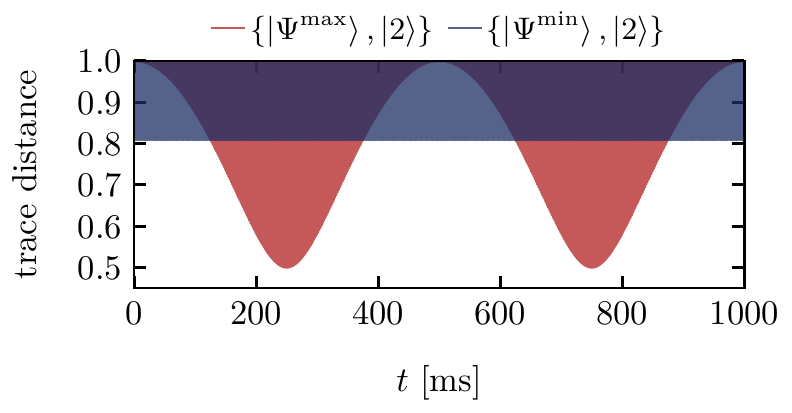}
  \caption{%
    Top: Evolution of the population in the Fock states with $n = 0, 1$ and $2$
    (red, blue and green, respectively) when starting initially in $\ket{2}$
    (solid lines), $\ket{\Psi^\text{max}} = \ket{0}$ (dashed) and
    $\ket{\Psi^\text{min}} = \frac{1}{\sqrt{2}}\left(\ket{0} + i \ket{1}\right)$
    (dotted). Note that the dotted lines for $n = 0$ and $1$ are constantly
    equal to $0.5$.  The background seems to be filled because the blue and
    green solid lines, corresponding to the populations in $n = 1$ and $2$ when
    starting in $\ket{2}$, oscillate rapidly (see inset). The vertical dashed
    lines in the inset shows the position of the measurements.
    Bottom: Evolution of the trace distance for the state pairs
    $\{\ket{\Psi^\text{max}},\ket{2}\}$ (red) and
    $\{\ket{\Psi^\text{min}},\ket{2}\}$ (blue), respectively. The filled areas
    are again due to the fast oscillation. The parameters are the same as in
    \figref{sfig:optimal_state_pair}.
  }
  \label{sfig:compare_state_pairs}
\end{figure}

Having reduced the optimization to an optimization over a two-dimensional space
to identify the second state of the optimal pair, we proceed by parametrizing
this state by its Bloch angles,
\begin{align*}
  \ket{\psi (\vartheta,\varphi)} =
  \cos\left( \tfrac{\vartheta}{2}\right) \ket{0}
  + \sin\left( \tfrac{\vartheta}{2}\right) e^{i \varphi}\ket{1}\,.
\end{align*}
Next, we calculate the BLP measure numerically for different initial state pairs
$\{\ket{\Psi_1} = \ket{\psi(\vartheta,\varphi)}, \ket{\Psi_2} = \ket{2}\}$. The
result is shown in \figref{sfig:optimal_state_pair}. Note that the absolute value
of the BLP measure has no physical meaning since it depends, amongst others but
most notably, on $T$.  It is apparent from \figref{sfig:optimal_state_pair} that
there are several optimal state pairs, namely all states on the black circle
with $\varphi=\{0,\pi\}$ and $\vartheta\in\left[0,\pi\right]$. The two states
with $\varphi=\pm\tfrac{\pi}{2}$ and $\vartheta=\tfrac{\pi}{2}$ have the lowest
BLP measure. This observation can be explained with the help of
\figref{sfig:compare_state_pairs}: When starting in the state
$\ket{\Psi^\text{max}} \equiv \ket{\psi (\vartheta=0,\varphi=0)}=\ket{0}$ which
is one of the optimal states, the population in the states $\ket{0}$ and
$\ket{1}$ oscillate with $\alpha$, cf.  the dashed lines in
\figref{sfig:compare_state_pairs} (top). When starting in the Zeno level
$\ket{2}$, the population oscillates quickly between $\ket{1}$ and $\ket{2}$ due
to the Zeno pulse (solid lines, see also the inset). The trace distance of this
state pair is maximal when the first state is in $\ket{0}$ because it has no
overlap with $\ket{1}$ or $\ket{2}$, seen e.g.\ at $\unit[0]{ms}$ in
\figref{sfig:compare_state_pairs} (red area in the lower plot). On the other
hand, the trace distance decreases to $0.5$ when the first state is in $\ket{1}$
because it is, averaged over the fast oscillation, overlapping with half of the
second state of the pair. Thus, the trace distance oscillates slowly with
$\alpha$ between $0.5$ and $1$, with a fast underlying oscillation with
$\omega_Z$. In contrast, $\ket{\Psi^\text{min}} \equiv \ket{\psi
(\vartheta=\tfrac{\pi}{2},\varphi=\tfrac{\pi}{2})}
= \frac{1}{\sqrt{2}}\left(\ket{0} + i \ket{1}\right)$, which leads to the
minimal BLP measure, behaves differently since it is an eigenstate of the
process in the Zeno limit, cf. dotted lines in \figref{sfig:compare_state_pairs}
(top).  Thus, half of the state overlaps with the $\ket{1}$-contribution when
starting in the Zeno level all the time and the trace distance is constant with
about $0.81$, with a fast underlying oscillation of the Zeno pulse, cf.
\figref{sfig:compare_state_pairs} (blue area in the lower plot). Since the red
area is larger than the blue one, the state $\ket{\Psi^\text{max}}$ leads to
a higher BLP measure than $\ket{\Psi^\text{min}}$.

These observations can be easily  generalized to all states on the black circle
in \figref{sfig:optimal_state_pair} since the cavity drive $S$, cf.\
\eqref{eq:H_CS} in the main text, induces a rotation around the $y$-axis of the
Bloch sphere for our choice of $\alpha$. Therefore, every initial state lying on
this circle will stay on it, leading to the same BLP measure for all of them. In
more general terms, the rotation axis $\vec{n}$ depends on the phase of $\alpha$
via
$\varphi_\alpha = \arccos\left(\frac{\text{Re}\,\alpha}{\abs{\alpha}}\right)$,
\begin{align*}
  \vec{n} = \left(\cos \varphi_\alpha,\, \sin \varphi_\alpha,\, 0\right)
          = \left(\text{Re}\,\alpha,\,
                  \text{Im}\,\alpha,\, 0\right)/\abs{\alpha}.
\end{align*}
The optimal state pair is obtained when choosing $\varphi = \varphi_\alpha
+ \pi/2$ for the state in the Zeno subspace.

When considering two Zeno pulses, $Z_1$ and $Z_2$, to induce non-Markovianity
and dissipation, the optimal state pair is possibly different since all three
states in the extended Zeno subspace $\mathcal{H}_{Z}^+$ take part in the
non-unitary dynamics induced by the Zeno coupling and the measurement.
A re-optimization was performed for the specific case of $\{\beta
= 0.025,\phi_2=4\pi,\phi_1=0.25\pi\}$ providing both dissipation ($S_L = 0.48$)
and a large amount of non-Markovianity (about twice as large as without having
the second Zeno pulse $Z_1$).  When optimizing $\ket{\Psi_1}$ as before, the
optimal state pair for this set of parameters was found to be
$\left\{\ket{\Psi_1} = \ket{\psi (\vartheta=0.56\pi,\varphi=1.92\pi)},\,
\ket{\Psi_2} = \ket{2}\right\}$.  To be exact, this optimization would have to
be performed for all considered parameters $\{\beta,\phi_2,\phi_1\}$ and with
considering choices other than $\ket{\Psi_1} = \ket{\psi (\vartheta,\varphi)}$
and $\ket{\Psi_2} = \ket{2}$ for the optimal state pair. However, the minimal
and maximal value of the BLP measure differ by about $1\%$ only and since the
absolute value of the BLP measure is not important, we choose
$\left\{\ket{\Psi_1} = \frac{1}{\sqrt{2}} \left( \ket{0} + \ket{1}\right),\,
\ket{\Psi_2} = \ket{2}\right\}$ throughout for simplicity.

\section{Derivation of the Lindblad master equation}

We now provide the detailed derivation of the Lindblad master equation
(\ref{eq:mastereq}) starting from the piecewise dynamics of
\eqref{eq:piecewise_dyn}. The first step is to evaluate the partial trace and to
derive the Krauss representation of the map.  To this end, we have to expand the
time evolution operator $\op{U}(dt)$ using the Baker-Campbell-Hausdorff formula
up to first order,
\begin{align*}
  \op{U}(dt)
  &= e^{-i\left(\op[CS]{H} + \op[AC]{H} + \op[Z]{H}\right)dt}\\
  &\approx e^{-i\op[Z]{H}dt}e^{-i\op[CS]{H}dt}e^{-i\op[AC]{H}dt}
\end{align*}
Using $\com{\op[AC]{H}}{\ket{h}\bra{h}} = 0$ and the fact that we start each
sequence with the initial state $\op{\rho}(t) =\op[C]{\rho}(t) \otimes
\ket{h}\bra{h}$, we can evaluate the partial trace explicitly. We arrive at
\begin{align}
  \op[C]{\rho}(t+dt)
  = \sum_{j=h,g,e} w_{jh} \, \op[CS]{U}(dt) \, \op[C]{\rho}(t) \,
                   \op[CS]{U}^\dagger(dt) \, w_{jh}^\dagger
  \label{seq:krauss}
\end{align}
with $\op[CS]{U}(dt) = e^{-i\op[CS]{H}dt}$ and $w_{jh}
= \bra{j}\op[Z]{U}(dt)\ket{h}$ with $\op[Z]{U}(dt) = e^{-i\op[Z]{H}dt}$.
Diagonalisation of $\op[Z]{U}(dt)$ leads to \cite{Raimond2012}
\begin{align*}
  \op[Z]{U}(dt)
    = e^{-i\tfrac{\phi_z}{2}} \op[+]{P}
    + e^{i\tfrac{\phi_z}{2}} \op[-]{P}
    + \op{P}_\perp
\end{align*}
with the Rabi angle $\phi_z = \omega_Z dt$, the projectors $\op{P}_\pm
= \ket{u_\pm}\bra{u_\pm}$, $\op{P}_\perp = \one - \op[+]{P} - \op[-]{P}$ and the
states $\ket{u_\pm} = \tfrac{1}{\sqrt{2}} (\ket{h,z} \pm \ket{+,z})$. This
expression can be used to evaluate the operators $w_{jh}$,
\begin{subequations}
  \label{seq:krauss_ops}
  \begin{eqnarray}
    w_{hh} &=& \op[C]{\mathds{1}} - \left(1 - \cos\tfrac{\phi_z}{2}\right)  \op{\Pi}\,,\\
    w_{gh} &=& \tfrac{1}{\sqrt{2}} \, \sin\tfrac{\phi_z}{2} \, \op{\Pi}\,, \\
    w_{eh} &=& \tfrac{1}{\sqrt{2}} \, \sin\tfrac{\phi_z}{2} \, \op{A}\,,
  \end{eqnarray}
\end{subequations}
where $\op{\Pi}$ and $\op A$ are photonic operators, $\op{\Pi} = \ket{z}\bra{z}$
and $\op{A} = \ket{z-1}\bra{z}$. Note that \eqref{seq:krauss} can be interpreted
both as the description of a measurement process with three measurement
operators and as the Krauss representation of the evolution of an open quantum
system.

Furthermore, a time-continuous master equation in Lindblad form can be derived
by assuming the measurement to be performed continuously with constant rate
$\kappa=\frac{1}{\tau}$ such that the number of measurements in a time interval
$dt$ is $\kappa dt$. This can be interpreted as ``smearing'' of one measurement
over the whole time interval $\tau$. As a result, the master equation will not
describe the same evolution as the piecewise dynamics for large values of
$\tau$. Using this assumption and going to the interaction picture with respect
to the coherent evolution $\op[CS]{H}$, we can rewrite \eqref{seq:krauss} to
\cite{Lucas2014}
\begin{align*}
  \op[C]{\rho}(t+dt) = \kappa \, dt \sum_{j=h,g,e}
    w_{jh} \, \op[C]{\rho} (t) \, w_{jh}^\dagger
    + \left( 1 - \kappa \, dt \right) \, \op[C]{\rho} (t)\,.
\end{align*}
Finally, we calculate the derivative of the reduced state as
$\frac{d\op[C]{\rho}(t)}{dt} = \lim_{dt\rightarrow 0} \frac{\op[C]{\rho}(t+dt)
- \op[C]{\rho}(t)}{dt}$ and go back to the non-interacting picture \cite{Lucas2014},
\begin{align*}
  \frac{d\op[C]{\rho}(t)}{dt} =
  &- i \com{\op[CS]{H}}{\op[C]{\rho}}\\
  &+ \kappa \left(
    \sum_{j=h,g,e}
    w_{jh} \, \op[C]{\rho} (t) \, w_{jh}^\dagger
  - \op[C]{\rho} (t) \right)\,.
\end{align*}
After inserting the Krauss operators of \eqref{seq:krauss_ops} and rearranging
the terms using that $\op{\Pi}$ is idempotent and $\op{\Pi}
= \op{A}^\dagger\op{A}$, we arrive at the master equation as shown in
\eqref{eq:mastereq}.

\begin{figure}[tb]
  \includegraphics{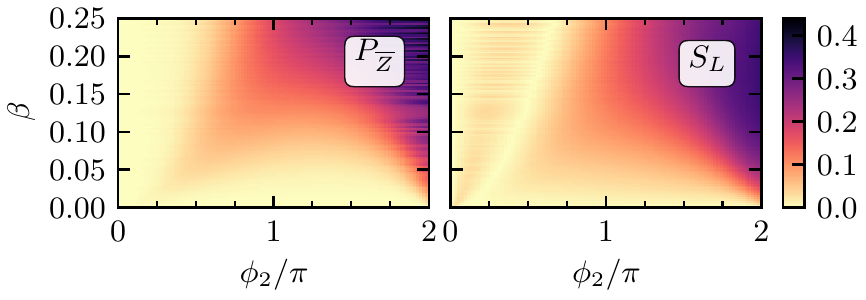}
  \caption{%
    Difference obtained when solving the master equation and using the
    piecewise dynamics for the Zeno infidelity $P_{\bar{Z}}$ (left) and the
    dissipation $S_L$ (right) as a function of the displacement $\beta$ due to
    the classical drive $S$ and the Rabi angle $\phi_2$ accumulated due to the
    Zeno pulse $Z_2$.
  }
  \label{sfig:markov_colormap}
\end{figure}

To assess the agreement of the master equation (\ref{eq:mastereq}) with the
piecewise dynamics of \eqref{eq:piecewise_dyn}, we propagate the initial state
$\ket{0}$ with $z=2$ using both methods. \figref{sfig:markov_colormap} shows the
difference in infidelity of the QZD, $P_{\bar{Z}}$, and in the linear entropy
$S_L$ between the two pictures. It can be seen that they show good agreement for
small Rabi angles $\phi_2$ or displacements $\beta$. However, the two pictures
deviate when our assumptions during the derivation of the master equation are
violated.  Firstly, the agreement deteriorates for large $\beta$, since this
contradicts the assumption of continuous measurements, $dt \rightarrow 0$.
Secondly, we have used the Baker-Campbell-Hausdorff formula up to first order to
derive the Krauss representation. We thus make an error which is of the order of
the commutator of $\op[\cs]{H}$ and $\op[z]{H}$ which scales as $\phi_z\beta$.
Therefore, the deviation also gets larger as $\phi_2$ increases.

\section{Effects leading to Zeno confinement}

Quantum Zeno dynamics describe the effect of confining the dynamics of a system
to a tailored subspace by frequent measurements. The dynamics shown
in~\figref{fig:nonmarkov_colormap} and~\ref{sfig:markov_colormap} agree with
this intuitive picture since the Zeno infidelity vanishes as the displacement
$\beta$, which is proportional to the time inbetween two measurements, goes to
zero. However, Zeno confinement occurs also in other regions as will be
explained in the following.

(i) QZD can not only be induced by frequent measurements but also by strongly
coupling the system to a meter. In fact, the two cases are formally equivalent
in the limit of a fast repetition rate and strong coupling \cite{Facchi2019}. In
our case, the strength of the coupling between the HO and the three-level system
is given by the field strength of the Zeno pulse, $\omega_z = \phi_z/\tau$,
which is proportional to the Rabi angle $\phi_z$.  Thus, the quality of the Zeno
confinement is very good for large values of the Rabi angle $\phi_z$ even if
$\beta$ is large.

(ii) The coupling of the Zeno level to the dressed state $\ket{+,z}$ induces
a decay of the Zeno level's population to the level below. This can be explained
by considering the piecewise dynamics of \eqref{eq:piecewise_dyn} and the effect
of the Zeno pulse $Z_z$.  If the initial state $\ket{h,z}$ is evolved under the
action of the Hamiltonian $\op{H}_z$ in \eqref{eq:H_z} (while neglecting the
action of the source $S$), the atom-field state at time $\tau$ is given by
\begin{align}
  \ket{\Psi_{\ket{h,z}}(\tau)} = \cos \left(\frac{\phi_z}{2}\right) \ket{h,z}
                               + \sin \left(\frac{\phi_z}{2}\right) \ket{+,z},
  \label{seq:rabi}
\end{align}
where $\phi_z = \omega_z \tau$ is the Rabi angle. All other states $\ket{h,n\neq
z}$ are unaffected by $\op{H}_z$.  It can be seen that, when performing a Zeno
pulse with Rabi angle $\phi_z \neq 2n\pi$, a fraction of
$\sin^2\left(\frac{\phi_z}{2}\right)$ of the population that has been in the
state $\ket{h,z}$ ends up in the state $\ket{+,z}$. According to the definition
of the dressed states, half of the photons are in the state $\ket{z-1}$ here,
i.e.\ back in the Zeno subspace, which enhances the Zeno confinement. In the
Lindblad master equation \ref{eq:mastereq}, the same effect is achieved by the
Lindblad operator $\op{A}$ as can be seen from the reappearing
$\sin^2\left(\frac{\phi_z}{2}\right)$-contribution in the decay rate $\gamma_A$.
As a consequence, if the initial state is lying above the Zeno level, $n_0 > z$,
the Zeno pulse has to be changed to be resonant to the $\ket{h,z}
\leftrightarrow\ket{+,z+1}$ transition, to also benefit from this enhancement.

(iii) For $\pi<\phi_z<2\pi$, the phase of the $\ket{h,z}$ contribution in
\eqref{seq:rabi} is negative due to the cosine. This leads to an inversion of
the dynamics back into the Zeno subspace which improves the QZD additionally.

\end{appendix}

\end{document}